\documentclass[twocolumn,preprintnumbers,amsmath,amssymb]{revtex4}

\usepackage{graphicx}
\usepackage{dcolumn}
\usepackage{bm}

\newcommand{\be}[1]{\begin{equation}\label{eq:#1}}
\newcommand{\ee}{\end{equation}}
\newcommand{\bea}{\begin{eqnarray}}
\newcommand{\eea}{\end{eqnarray}}

\newcommand{\bt}{\textbf}

\newcommand{\phd}{\phantom{\dag}}
\newcommand{\ph}{\phantom{.}}
\newcommand{\up}{^{\phd}}
\newcommand{\noi}{\noindent}
\newcommand{\no}{\nonumber}

\newcommand{\dv}[1]{\partial_{#1}\up}
\newcommand{\dvb}[1]{\bm{\nabla}_{\bm{#1}}\up}

\begin{document}
\def\v#1{{\bf #1}}


\title{Berry phase and topological spin transport in the chiral d-density wave state}

\author{P. Kotetes}\email{pkotetes@central.ntua.gr}
\author{G. Varelogiannis}
\affiliation{Department of Physics, National Technical University
of Athens, GR-15780 Athens, Greece}



\maketitle


\noi\bt{Abstract} In this paper we demonstrate the possibility of
dissipationless spin transport in the chiral d-density wave state,
by the sole application of a uniform Zeeman field gradient. The
occurrence of these spontaneous spin currents is attributed to the
parity (${\cal P}$) and time-reversal (${\cal T}$) violation
induced by the $d_{xy}+id_{x^2-y^2}$ density wave order parameter.
We calculate the spin Hall conductance and reveal its intimate
relation to the Berry phase which is generated when the Zeeman
field is applied adiabatically. Finally, we demonstrate that in
the zero temperature and doping case, the spin Hall conductance is
quantized as it becomes a topological invariant.\\

\noi\bt{Keywords} Berry phase $\cdot$ Spontaneous spin Hall effect $\cdot$ chiral d-density wave\\

\noi\bt{PACS} 73.43.-f $\cdot$ 72.25.-b $\cdot$ 71.27.+a \\

\noi\bt{1 Introduction}\\

\noi Manipulating spin currents via the application of electric
fields has received notable attention lately, especially in the
field of semiconductors \cite{ZhangScience,ExpSpin}. In these
materials the key ingredient for spin transport is the presence of
strong spin-orbit coupling. The arising spin currents may be of
intrinsic or extrinsic origin. In both cases, generating spin
currents based on spin-orbit coupling is inherently pathological.
The spin-orbit term is essentially a spin non-conserving
interaction term. As a consequence, the absence of spin
conservation prevents us from defining a proper spin current.
Nevertheless, Murakami et al, have managed to define
dissipationless spin currents, showing in addition, that the
related spin Hall conductance depends solely on the curvature of
an $SU(2)$-holonomy \cite{ZhangPRB}. Actually, this treatment
focuses on how to cast out the dissipative part of the spin
currents. In another work \cite{SpinHelix}, it has been shown that
the interplay of Rashba and Dresselhaus spin-orbit terms can lead
to perfectly dissipationless spin currents. However, the latter
case is extremely sensitive to the parameters of the spin-orbit
couplings.

In this direction, we propose an alternative way of generating
dissipationless spin currents. Specifically, we consider the case
of the chiral d-density wave which is an unconventional spin
singlet long-range order. Recently the chiral d-density wave state
was proposed as a candidate state \cite{LT,Meissner}, which can
simultaneously explain the enhanced diamagnetic and Nernst signals
observed in the pseudogap regime of the cuprates \cite{Exp Nerst
Magnet}. Moreover, it has also been considered \cite{Tewari} to be
the origin of time reversal breaking in YBCO \cite{Xia}. Apart
from the prominent significance of this state in understanding
high Tc superconductivity, the chiral d-density wave constitutes a
paradigm of a functional state of matter. It has already been
shown to exhibit the dissipationless electric charge Hall
transport by the sole application of an electric field
\cite{KVL,Yakovenko}. In fact as we shall demonstrate the
topological spin Hall transport is simply the spin analogue of
dissipationless charge transport, with the gradient of the Zeeman
field playing the role of the electric field
\cite{SpinHall,Volovik}.

In the case of the chiral d-density wave state, the $SU(2)$ spin
rotational symmetry is totally preserved, which leads to a
straightforward definition of a conserved spin current. The
dissipationless character of spin Hall transport resides solely on
the unconventional structure of the order parameter of the density
wave. The chiral d-density wave is a state of two coexisting
d-wave type density waves. The $d_{xy}$ component violates parity
(${\cal P}$) in two dimensions and the $d_{x^2-y^2}$ component
violates time-reversal (${\cal T}$). As a result, an anomalous
magneto-electric coupling arises in the effective action of the
charge $U(1)$ and spin $SU(2)$ gauge fields, leading to a
spontaneous charge and spin Hall response (See e.g.
\cite{Volovik}). From another point of view, the adiabatic
application of external fields gives rise to a $U(1)$ holonomy in
$\bm{k}-$space and a concomitant Berry connection \cite{Berry},
that provides an anomalous velocity \cite{semiclassical} to the
quasi-particles of the system. In fact, the latter situation is
similar to a ferromagnetic system \cite{LuttingerFerromagnetics}.
However, in our case there is a finite \textit{orbital}
magnetization \cite{OrbitalMagnetization}
in contrast to the usual magnetization.\\

\noi\bt{2 The Model}\\

\noi To demonstrate how the Spontaneous Spin Hall effect arises,
we shall consider the chiral d-density wave Hamiltonian ${\cal
H}=\sum_{\bm{k}}\Psi_{\bm{k}}^{\dag}{\cal
H}(\bm{k})\Psi_{\bm{k}}\up$ with \bea{\cal
H}(\bm{k})=\left(\begin{array}{cc}
\delta(\bm{k})+g_3\up(\bm{k})&g_{1}\up(\bm{k})-ig_2\up(\bm{k})\\
g_{1}\up(\bm{k})+ig_2\up(\bm{k})&\delta(\bm{k})-g_3\up(\bm{k})\\\end{array}\right),\eea

\noi where we have introduced the isospinor
$\Psi_{\bm{k}}^{\dag}=(c_{\bm{k}}^{\dag}\ph c_{\bm{k+Q}}^{\dag})$
and restricted the $\bm{k}-$space summation in the
\textit{reduced} Brillouin zone. The operators
$c_{\bm{k}}\up/c_{\bm{k}}^{\dag}$ annihilate/create an electron of
momentum $\bm{k}$. For simplicity we have omitted the spin
indices, which will be suitably embodied in our formalism later.
Moreover, we employ a single band Bloch electron model with
particle-hole asymmetric and symmetric kinetic terms
$\delta(\bm{k})=4t'\cos k_x\cos k_y-\mu$ and $g_3(\bm{k})=-2t(\cos
k_x+\cos k_y)$. We have also introduced the $d_{xy}$ and
$d_{x^2-y^2}$ order parameters, $g_1(\bm{k})=\Delta_{1}\sin
k_x\sin k_y$ and $g_2(\bm{k})=-\Delta_{2}(\cos k_x-\cos k_y)$,
respectively. It is straightforward to observe the similarity to
the ferromagnetic case by writing this Hamiltonian in the
equivalent form ${\cal
H}(\bm{k})=\delta(\bm{k})+E(\bm{k})\hat{\bm{g}}(\bm{k})\cdot\bm{\tau}$
with $\hat{\bm{g}}(\bm{k})=\bm{g}(\bm{k})/|\bm{g}(\bm{k})|$ and
$\bm{\tau}$ the isospin Pauli matrices. It is the existence of an
isospin vector which can be rotated in isospin space under the
adiabatic application of an external field, that generates a Berry
phase responsible for the non-dissipative charge and spin Hall
response.\\

\noi\bt{3 Application of an electric field}\\

\noi Our next step is to perturb our system with an external
electric field and calculate the emergent Berry curvature. This
may be effected by considering a time-dependent vector potential
of the form $\bm{A}(t)=-\bm{{\cal E}}t$. This type of perturbation
enters our Hamiltonian via the Peierls-Onsager substitution
$\bm{k}\rightarrow\bm{k}+e\bm{A}(t)$ with $e>0$. As a consequence,
this minimal coupling constitutes the system's Hamiltonian
time-dependent, ${\cal H}(\bm{k})\rightarrow{\cal H}(\bm{k},t)$.
When this parameter changes adiabatically along a closed path, the
ground state of the system remains unaltered and as an outcome the
wavefunction of the system acquires a Berry phase. As we have
already pointed out, the generation of the Berry phase is strongly
related to the ${\cal P}-{\cal T}$ violation originating from the
chiral character of the density wave and permits the
dissipationless charge and spin pumping.

When the external perturbation is present, the exact eigenstates
of the system, $|\Psi_{\nu}(\bm{k},t)\rangle$, satisfy the
parametric Schr\"{o}dinger equation ${\cal
H}(\bm{k},t)|\Psi_{\nu}\up(\bm{k},t)\rangle=i\dv{t}|\Psi_{\nu}\up(\bm{k},t)\rangle$.
In the adiabatic approximation, any exact eigenstate of the
parametric Hamiltonian, is considered to acquire only a phase
factor. This phase factor can be dissociated into two parts, the
dynamical part and the Berry phase. Equivalently, one considers
the following form for the exact eigenstates
\bea|\Psi_{\nu}\up(\bm{k},t)\rangle=e^{-i\int_{0}^tE_{\nu}\up(\bm{k},t'){\rm
dt'}+i\gamma_{\nu}\up(\bm{k},t)}|\Phi_{\nu}\up(\bm{k},t)\rangle\,,\eea

\noi where the first part of the phase denotes the acquired
dynamical phase and the second, the Berry phase
$\gamma_{\nu}(\bm{k},t)$. The states $|\Phi_{\nu}(\bm{k},t)$, are
the instantaneous (snapshot) eigenstates of the parametric
Hamiltonian, satisfying ${\cal
H}(\bm{k},t)|\Phi_{\nu}(\bm{k},t)\rangle=
E_{\nu}(\bm{k},t)|\Phi_{\nu}(\bm{k},t)\rangle$. In this equation,
$t$, is introduced only as a parameter. This means that these
snapshot eigenstates are not really time-dependent but only
parameter dependent, which in our case coincides with $t$. In our
case we deal with a two band system, characterized by the snapshot
eigenstates $|\Phi_{\pm}\rangle$ and the corresponding
eigenenergies $E_{\pm}(\bm{k},t)=\delta(\bm{k},t)\pm E(\bm{k},t)$.

The definition of the exact eigenstates $|\Psi_{\nu}\rangle$ in
terms of the snapshot eigenstates $|\Phi_{\nu}\rangle$ readily
provides the time dependence of the Berry phase
\bea\gamma_{\nu}\up(\bm{k},t)-\gamma_{\nu}\up(\bm{k},0)
=\int_0^t{\rm dt'}\ph\langle\Phi_{\nu}\up(\bm{k},t')|
i\dv{t'}|\Phi_{\nu}\up(\bm{k},t')\rangle\,.\eea

\noi Setting $\gamma_{\nu}\up(\bm{k},t)=0$ and taking into account
that the time dependence of the snapshot eigenstates arises
because the crystal momentum becomes time dependent,
$\bm{k}\rightarrow\bm{k}(t)=\bm{k}-e\bm{{\cal E}}t$, yields
$\Gamma_{\nu}\up(\bm{{\cal E}})=$$\int_0^T{\rm dt'}\ph
\langle\Phi_{\nu}\up(\bm{k}(t'))|
i\dv{t'}|\Phi_{\nu}\up(\bm{k}(t'))\rangle$$=\oint_{C(\bm{{\cal
E}})}{\rm d}\bm{{\rm k}}\cdot\ph\langle\Phi_{\nu}\up(\bm{k})|
i\dvb{k}|\Phi_{\nu}\up(\bm{k})\rangle=\oint_{C(\bm{{\cal E}})}{\rm
d}\bm{{\rm k}}\cdot\bm{{\cal
A}}_{\nu}\up(\bm{k})=\int_{S(\bm{{\cal E}})}{\rm d^2k}\ph
\Omega_{\nu}^z(\bm{k})$, where $\Gamma_{\nu}$ is the total Berry
phase generated for a period $T$ of the adiabatic pumping
\cite{Zak}. We have also introduced the $U(1)$ Berry connection
$\bm{{\cal A}}_{\nu}\up(\bm{k})=\langle\Phi_{\nu}\up(\bm{k})\mid
i\dvb{k}\mid\Phi_{\nu}\up(\bm{k})\rangle$ and the Berry curvature
$\bm{\Omega}_{\nu}\up(\bm{k})=\dvb{k}\times\bm{{\cal
A}}_{\nu}\up(\bm{k}) =\Omega^z_{\nu}\ph\bm{\hat{z}}$. We observe
that the two-dimensional character of our system, forces the Berry
curvature to lie along the z-axis.

By defining
$g_1\up(\bm{k})=E(\bm{k})\sin\theta(\bm{k})\cos\varphi(\bm{k})$,
$g_2\up(\bm{k})=E(\bm{k})\sin\theta(\bm{k})\sin\varphi(\bm{k})$
and $g_3\up(\bm{k})=E(\bm{k})\cos\theta(\bm{k})$, we obtain a
convenient expression for the snapshot eigenstates of the system
\bea|\Phi_{+}\up(\bm{k},t)\rangle=\left(\begin{array}{l}\cos\left(\frac{\theta(\bm{k},t)}{2}\right),\phantom{-}
\sin\left(\frac{\theta(\bm{k},t)}{2}\right) e^{i\varphi(\bm{k},t)}\end{array}\right)^T,&\\\no\\
|\Phi_{-}\up(\bm{k},t)\rangle=\left(\begin{array}{l}\sin\left(\frac{\theta(\bm{k},t)}{2}\right),
-\cos\left(\frac{\theta(\bm{k},t)}{2}\right)
e^{i\varphi(\bm{k},t)}\end{array}\right)^T,&\eea

\noi where $T$ denotes matrix transposition. The lower band's
Berry curvature is written \cite{C Zhang}
\bea\Omega_-^z(\bm{k})=-\frac{1}{2E^3(\bm{k})}\ph\bm{g}(\bm{k})\cdot\left(\frac{\partial\bm{g}(\bm{k})}{\partial
k_x\up}\times\frac{\partial\bm{g}(\bm{k})}{\partial
k_y\up}\right).\eea

\noi We observe that if time reversal is not violated, we have
$g_2(\bm{k})=0$, and the Berry curvature is zero. On the other
hand, if parity in two-dimensions is preserved then the Berry
curvature is non zero but its integral is. So \textit{no} net flux
of the $U(1)$ gauge field arises and \textit{no} anomalous
transport is permitted. The same stands for the Berry curvature of
the upper band, as we have
$\Omega_+^z(\bm{k})=-\Omega_-^z(\bm{k})$. Obviously, if the two
bands are equally occupied then the total Berry curvature is
zero.\\

\noi\bt{4 Dissipationless charge transport}\\

\noi Having obtained the expression for the Berry curvature we may
proceed with studying the dissipationless transport of the system.
As a warmup we shall study first the case of dissipationless
charge transport. As we shall demonstrate in the next section, the
Spontaneous Spin Hall effect will arise as a direct consequence of
the Spontaneous Charge Hall effect and the existence of gauge
invariance.

In order to demonstrate how the Spontaneous Charge Hall effect
arises, we have to define a charge current. This is easily
achieved by considering the equation of continuity of the electric
charge in momentum space, which dictates that
$\dot{\rho}_c\up(\bm{q})+i\bm{q}\cdot\bm{J}_c\up(\bm{q})=0\Rightarrow
\bm{J}_c\up(\bm{q})=i\lim_{\bm{q}\rightarrow
0}\bm{\nabla_{q}}\ph\dot{\rho}_c\up(\bm{q})$. The charge density
$\rho_c\up(\bm{q},t)$ can be expressed using the snapshot
eigenstates as $\rho_c\up(\bm{q},t)=-e\sum_{\bm{k},\nu=\pm}
\langle\Phi_{\nu}\up(\bm{k+q},t)\mid\Phi_{\nu}\up(\bm{k},t)\rangle$.
To obtain the charge current we need the time derivative of the
charge density which can be written as
\bea\dot{\rho}_c\up(\bm{q},t)
&=&-e\sum_{\bm{k},\nu=\pm}\left\{\langle\dv{t}\Phi_{\nu}\up(\bm{k},t)|\Phi_{\nu}\up(\bm{k}-\bm{q},t)\rangle\right.\no\\
&&\left.\qquad\quad+\langle\Phi_{\nu}\up(\bm{k}+\bm{q},t)|\dv{t}\Phi_{\nu}\up(\bm{k},t)\rangle\right\}\,,\eea

\noi where we have used that
$\langle\dv{t}\Phi_{\nu}\up(\bm{k+q},t)\mid\Phi_{\nu}\up(\bm{k},t)\rangle=
\langle\dv{t}\Phi_{\nu}\up(\bm{k},t)\mid\Phi_{\nu}\up(\bm{k}-\bm{q},t)\rangle$.
Under these conditions the charge current becomes
\bea\bm{J}_c\up&=&-ie\lim_{\bm{q}\rightarrow 0}\bm{\nabla_{q}}\ph
\sum_{\bm{k},\nu=\pm}\left\{\langle\dv{t}\Phi_{\nu}\up(\bm{k},t)\mid\Phi_{\nu}\up(\bm{k-q},t)\rangle\right.\no\\
&&\left.\qquad\qquad\qquad\qquad+
\langle\Phi_{\nu}\up(\bm{k+q},t)\mid\dv{t}\Phi_{\nu}\up(\bm{k},t)\rangle\right\}\no\\
&=&ie\sum_{\bm{k},\nu=\pm}\left\{\langle\dv{t}\Phi_{\nu}\up(\bm{k},t)\mid\dvb{k}\Phi_{\nu}\up(\bm{k},t)\rangle-h.c.\right\}\no\\
&=&-ie^2{\cal E}_i\up\sum_{\bm{k},\nu=\pm}\sum_{i=x,y}
\left\{\langle\dv{k_i}\Phi_{\nu}\up(\bm{k})\mid\dvb{k}\Phi_{\nu}\up(\bm{k})\rangle-h.c.\right\}\no\\
&=&e^2(\bm{{\cal
E}}\times\bm{\hat{z}})\sum_{{\nu=\pm}}\int_{RBZ}\frac{{\rm
d^2k}}{(2\pi)^2}\ph
n_F\up[E_{\nu}\up(\bm{k})]\ph\Omega_{\nu}^z(\bm{k})\,,\label{eq:Jc}\eea

\begin{figure}[t]\centering
\includegraphics[width=0.45\textwidth]{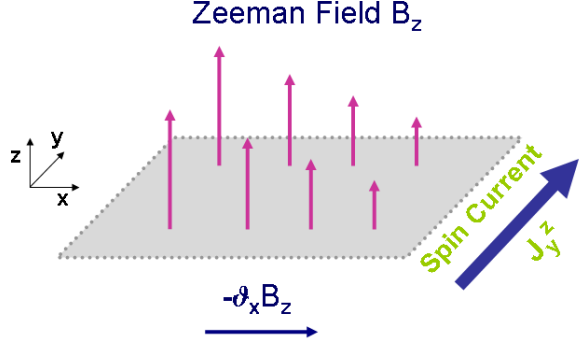}
\caption{{Spontaneous Spin Hall Effect setup. A Zeeman field
gradient leads to a dissipationless spin Hall response. The
polarization of the spin current is parallel to the external
magnetic field. In the zero temperature and doping regime the spin
Hall conductance is quantized.}}\label{fig:SSQHE}
\end{figure}

\noi where $n_F\up$ is the Fermi-Dirac distribution. At zero
temperature, if the lower and upper bands are separated by a gap,
only the lower band is occupied and we obtain
$n_F\up[E_-\up(\bm{k})]=1$ and $n_F\up[E_+\up(\bm{k})]=0$. In this
case the charge current becomes
$\bm{J}_c\up=-\frac{e^2}{2\pi}\widehat{N}(\bm{{\cal
E}}\times\bm{\hat{z}})$, where we have introduced the topological
invariant \cite{Volovik,TopologicalInvariant}
\bea\widehat{N}&=&-\frac{1}{2\pi}\int_{RBZ}{\rm
d^2k}\ph\Omega_{-}^z(\bm{k})\no\\
&=&\frac{1}{4\pi}\int_{RBZ}{\rm d^2k}\ph
\bm{\hat{g}}(\bm{k})\cdot\left(\frac{\partial\bm{\hat{g}}(\bm{k})}{\partial
k_x\up}\times\frac{\partial\bm{\hat{g}}(\bm{k})}{\partial
k_y\up}\right)\,,\eea \noi which is a winding number related to
the mapping of the reduced Brillouin zone to the order parameter
space. This winding number is an integer, as it corresponds to the
mapping of a torus $T^2$ to an $S^2$ sphere. In our case we obtain
$\widehat{N}=1$. This integer equals the angular momentum, in
$\bm{k}-$space, of the chiral ground state. One would expect that
$\widehat{N}=2$, as the order parameter is composed by d-wave
functions. However, this contribution is halved as we are
restricted to the reduced Brillouin zone.

It is straightforward to obtain the value of the Hall conductance
for the dissipationless charge transport from its defining
expression $\bm{J}_c\up=\sigma_{xy}^c(\bm{{\cal
E}}\times\bm{\hat{z}})$. We observe that the value of the Hall
conductance \cite{KVL} is universal and equal to (per one spin
component) \bea\sigma_{xy}^c=-\frac{e^2}{2\pi}=-\frac{e^2}{h}.\eea

\noi We have to underline that this universality originates from
the gauge invariance of our system. This should be contrasted to
the case of chiral superconductors where the Hall conductance is
affected by the existence of the Goldstone mode of the broken
$U(1)$ gauge invariance\cite{Goryo}.\\

\begin{figure}[t]\centering
\includegraphics[width=0.5\textwidth]{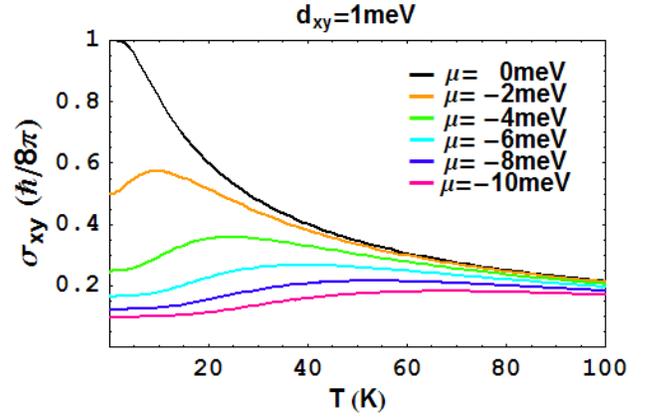}
\caption{{Temperature and doping dependence of the spin Hall
conductance, $\sigma_{xy}^s$. At zero temperature and chemical
potential the conductance is equal to the spin quantum
$\hbar/8\pi$. The increase of temperature or doping breaks down
the topological robustness of the chiral d-density wave,
decreasing the conductance. ($t=250meV$, $t'=0$,
$d_{x^2-y^2}=50meV$)}}\label{fig:Hallcond}
\end{figure}

\noi\bt{5 Dissipationless spin Hall transport}\\

\noi To demonstrate how the Spontaneous Spin Hall effect arises we
could embody in our formalism the electron spin and calculate the
spin Hall conductance in a manner similar to the previous section.
However, we shall follow another derivation that takes advantage
of gauge invariance and exhibits the intimate connection of
dissipationless charge and spin transport. Specifically, we would
like to show that the generation of dissipationless spin currents
by applying an external uniform Zeeman field gradient can be
mapped to the case of pumping charge with an external electric
field. We start from a general real space representation of our
chiral d-density wave Hamiltonian  \bea S&=&\int{\rm
dtd^2}x\ph\Psi_{\uparrow}^{\dag}(t,\bm{x})\left\{i\frac{\partial}{\partial
t}-{\cal H}(\bm{\hat{p}},\bm{x})\right\}
\Psi_{\uparrow}\up(t,\bm{x})\no\\
&+&\int{\rm dtd^2}x\ph\Psi_{\uparrow}^{\dag}(t,\bm{x})
\mu_z\up(\bm{x}\cdot\bm{\nabla}B_z\up)
\Psi_{\uparrow}\up(t,\bm{x})\,,\eea

\noi where $B_z$ is the external field and $\mu_z\up$ the magnetic
moment of the electron. Based on the fact that the chiral
d-density wave state is characterized by an $SU(2)$ spin
invariance we have considered one spin component. The Zeeman
interaction term can be eliminated by a phase transformation
$\Psi_{\uparrow}\up(t,\bm{x})\rightarrow
e^{i\varphi(x)}\Psi_{\uparrow}\up(t,\bm{x})$  with
$\varphi(x)=\mu_z\up(\bm{x}\cdot\bm{\nabla}B_z\up)t$. This
corresponds to a $U(1)$ gauge transformation, generating a vector
potential $\bm{A}_z(t)=\bm{\nabla}B_z\up t$. Consequently a
constant gradient of a Zeeman field corresponds to an electric
field $\bm{{\cal E}}=-\bm{\nabla}B_z$ and according to
Eq.(\ref{eq:Jc}), leads to the spin current
$\bm{J}_s\up=\sigma_{xy}^s(\bm{\hat{z}}\times\bm{\nabla}B_z\up)$.
If temperature and doping are zero the spin Hall conductance is
quantized $\sigma_{xy}^s=\mu^2_z/2\pi$. By substituting
$\mu_z=\hbar/2$, we obtain (per one spin component)
\bea\sigma_{xy}^s=\frac{\hbar}{8\pi}.\eea

In the general case, the spin Hall conductance is strongly
affected by temperature and doping. As we may see in
Fig.(\ref{fig:Hallcond}), both of them tend to suppress the
Spontaneous Spin Hall effect. If we take into account that the
robustness of dissipationless spin transport is directly related
to the robustness of the system's gap it is natural to expect such
a behaviour. A temperature bath gives rise to thermally excited
carries that occupy the upper band of our system which has the
opposite Berry curvature of the lower band. The same effect is
reproduced by a non zero chemical potential but in a different
manner. If the chemical potential crosses the upper band, both
bands become occupied yielding conflicting contributions of Berry
curvature. However, despite the negative effect that temperature
and doping have on dissipationless spin transport they may
constitute some of its controlling external parameters.\\

\noi\bt{6 Discussion}\\

\noi Motivated by the necessity of inquiring systems supporting
dissipationless spin currents, we have studied the occurrence of
topological spin transport in the chiral d-density wave state. The
chiral d-density wave is an unconventional spin-singlet state
giving rise to a Berry connection in $\bm{k}-$space. We have
demonstrated that it exhibits the Spontaneous Spin Hall effect
under the adiabatic application of an external Zeeman field
gradient. Specifically, spin transport is dissipationless and in
the zero temperature and doping case the spin Hall conductance
becomes a topological invariant and is quantized. In the general
case, spin transport is controlled by the presence of temperature
and chemical potential.\\

\noi\bt{\textit{Acknowledgments}} The authors are grateful to
Professor P. B. Littlewood for enlightening discussions. This work
has been supported by the EU STRP grant NMP4-CT-2005-517039. P.K.
also acknowledges financial support by the Greek Scholarships
State Foundation.


\begin{thebibliography}{00}

\bibitem{ZhangScience} S. Murakami, N. Nagaosa, and S. C. Zhang, Science \bt{301}, 1348
(2003); J. Sinova, D. Culcer, Q. Niu, N. A. Sinitsyn, T.
Jungwirth, and A. H. MacDonald, Phys. Rev. Lett. \bt{92}, 126603
(2004); H. A. Engel, B. I. Halperin and E. I. Rashba, Phys. Rev.
Lett. \bt{95}, 166605 (2005).

\bibitem{ExpSpin}Y. K. Kato, R. C. Myers, A. C. Gossard, and D. D. Awschalom,
Science \bt{306}, 1910 (2004); J. Wunderlich, B. Kaestner, J.
Sinova, and T. Jungwirth, Phys. Rev. Lett. \bt{94}, 047204 (2005).

\bibitem{ZhangPRB}S. Murakami, N. Nagaosa, and S. C. Zhang, Phys. Rev. B \bt{69},
235206 (2004).

\bibitem{SpinHelix}B. A. Bernevig, J. Orenstein and S. C. Zhang, Phys. Rev. Lett. \bt{97}, 236601
(2006).


\bibitem{LT}P. Kotetes and G. Varelogiannis, work to appear for
the proceedings of the 25th LT conference.

\bibitem{Meissner}P. Kotetes and G. Varelogiannis, Phys. Rev. B., \bt{78}, 220509(R) (2008).

\bibitem{Exp Nerst Magnet} Yayu Wang et al, Phys. Rev. B \bt{73}, 024510
(2006); Yayu Wang et al, Phys. Rev. Lett. \bt{95}, 247002 (2005).

\bibitem{Tewari} S. Tewari et al, Phys. Rev. Lett. \bt{100}, 217004 (2008) .

\bibitem{Xia} J. Xia et al., Phys. Rev. Lett. \bt{100}, 127002 (2008).

\bibitem{KVL}P. Kotetes and G. Varelogiannis, Europhys. Lett., \bt{84}, 37012 (2008).

\bibitem{Yakovenko}V. M. Yakovenko, Phys. Rev. Lett. \bt{65}, 251 (1990).

\bibitem{SpinHall}G. E. Volovik and V. M. Yakovenko, J. Phys. Condens.
Matter \bt{1}, 5263 (1989); J. Goryo, Phys. Rev. B \bt{77}, 144504
(2008).

\bibitem{Volovik} G. E. Volovik, \textit{The Universe in a Helium Droplet}, Oxford Science Publications (2003).



\bibitem{Berry}M.V. Berry, Proc. R. Soc. London A \bt{392}, 45
(1984); R. Resta, Rev. of Mod. Phys. \bt{66}, 899 (1994).

\bibitem{semiclassical}M.-C. Chang and Q. Niu, Phys. Rev. B \bt{53}, 7010
(1996); G. Sundaram and Q. Niu, Phys. Rev. B \bt{59}, 14915
(1999).

\bibitem{LuttingerFerromagnetics}R. Karplus and J.M. Luttinger, Phys. Rev. \bt{95}, 1154
(1954); J.M. Luttinger, Phys. Rev. \bt{112}, 739 (1958); T.
Jungwirth, Q. Niu, and A. H. MacDonald, Phys. Rev. Lett. \bt{88},
207208 (2002); M. Onoda and N. Nagaosa, J. Phys. Soc. Jpn.
\bt{71}, 19 (2002).

\bibitem{OrbitalMagnetization}J. Shi, G. Vignale, D. Xiao and Q.
Niu, Phys. Rev. Lett. \bt{99}, 197202 (2007); T. Thonhauser, D.
Ceresoli, D. Vanderbilt, and R. Resta, Phys. Rev. Lett. \bt{95},
137205 (2005).

\bibitem{Zak}J. Zak, Phys. Rev. Lett. \bt{62}, 2747, (1989).

\bibitem{C Zhang}C. Zhang, S. Tewari, V. M. Yakovenko, and S. Das
Sarma, Phys. Rev. B., \bt{78}, 174508 (2008).

\bibitem{TopologicalInvariant}D. J. Thouless, M. Kohmoto, M. P.
Nightingale, and M. den Nijs, Phys. Rev. Lett. \bt{49}, 405
(1982).

\bibitem{Goryo}J. Goryo and K. Ishikawa, Phys. Lett. A \bt{260}, 294 (1999);
J. Goryo, Phys. Rev. B \bt{78}, 060501 (2008).


\end{thebibliography}
\end{document}